\documentclass[a4paper,10pt,twocolumn]{article}
\usepackage{tipa}
\usepackage[T1]{fontenc}
\usepackage{caption}
\usepackage{amsmath,amssymb}
\usepackage{comment}
\usepackage{adjustbox}
\usepackage{natbib}
\def\etal{\mbox{et al.}}

\usepackage{geometry}
\title{\huge Spectral Study of the Vocal Tract in Vowel Synthesis: \\ \vspace{8pt} \LARGE A Comparison between 1D and 3D Acoustic Analysis\vspace{8pt}} 
\author{Negar M. Harandi\ensuremath{^{\star}}, Daniel Aalto\ensuremath{^{\circ}}, Antti Hannukainen\ensuremath{^{\dagger}}, Jarmo Malinen\ensuremath{^{\dagger}}, Sidney Fels\ensuremath{^{\star}}}

\date{}

\begin{document}
\setlength{\abovedisplayskip}{10pt}
\setlength{\belowdisplayskip}{10pt}
\setlength{\abovedisplayshortskip}{10pt}
\setlength{\belowdisplayshortskip}{10pt}

\twocolumn[
  \begin{@twocolumnfalse}
  \maketitle
  \vspace{-10pt}
  \centering
   \ensuremath{^{\star}}University of British Columbia, Canada. \ensuremath{^{\circ}}University of Alberta, Canada. \ensuremath{^{\dagger}}Aalto University, Finland.
   
  \vspace{30pt}
  \end{@twocolumnfalse}
]

\begin{abstract}
   
\noindent A state-of-the-art 1D acoustic synthesizer has been previously developed, and coupled to speaker-specific biomechanical models of oropharynx in ArtiSynth. As expected, the formant frequencies of the synthesized vowel sounds were shown to be different from those of the recorded audio. Such discrepancy was hypothesized to be due to the simplified geometry of the vocal tract model as well as the one dimensional implementation of Navier-Stokes equations. In this paper, we calculate Helmholtz resonances of our vocal tract geometries using 3D finite element method (FEM), and compare them with the formant frequencies obtained from the 1D method and audio. We hope such comparison helps with clarifying the limitations of our current models and/or speech synthesizer.

\end{abstract}

\section{Introduction}
Articulatory speech synthesisers generate sound based on the shape of the vocal tract. Vibration of the vocal folds under the expiratory air flow is the source in the system; and the vocal tract, consisting of the larynx, pharynx, oral and nasal cavities, constitutes a filter where sound frequencies are shaped. This creates a number of resonant peaks in the spectrum, known as formants. The first and second formants ($F_{1}$ and $F_{2}$) are used to distinguish the vowel phonemes, where the value of $F_{1}$ and $F_{2}$ is controlled by the height and backness-frontness of the tongue body respectively. 

Traditionally, the acoustic system is approximated by a one-dimensional wave equation that associates the slow varying cross-sectional area of a rigid tube to the pressure wave for a low-frequency sound. However, complex shape of the vocal tract, with its side branches and asymmetry, has motivated higher dimensional acoustic analysis. The 3D analysis methods were shown to produce a better representation of the sound spectrum at the price of higher computational cost. However, some studies suggested that the spectrum yielded by 1D acoustic analysis matches closely that of the 3D analysis for frequencies less than 7KHz \citep{Takemoto2014, Arnela2014}. \cite{Aalto2012} suggested that the discrepancy between the resonance frequencies computed by 3D analysis of the vocal tract and the formant frequencies of the recorded audio is a result of insufficient boundary conditions in the wave equation especially in case of the open lips and/or velar port.

In this paper, we follow \cite{Aalto2014} in calculating the Helmholtz resonances of our vocal tract geometries using 3D FEM analysis. The resonances are then compared to the formant frequencies obtained from the 1D acoustic synthesizer proposed by \cite{Doel2008} and those of the recorded audio. 

\section{Material and Methods}

\begin{figure}[b!]
 \centering
 \captionsetup{font=small}
  \includegraphics[width=0.48\textwidth,keepaspectratio]{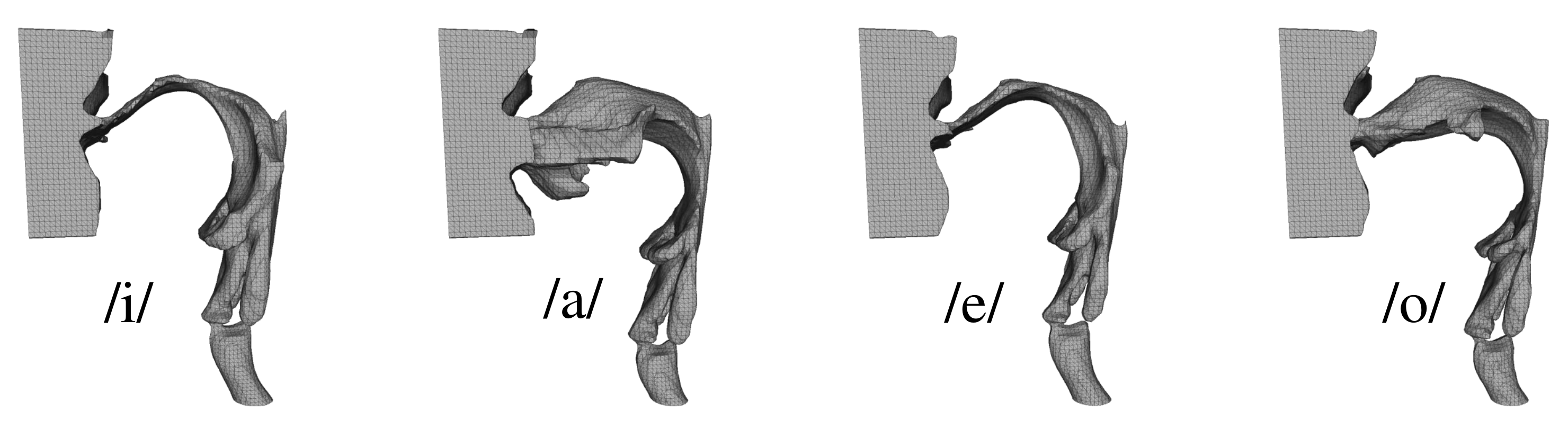}
  \caption [VT geometries extracted from MRI data] {VT geometries extracted from MRI data \citep{Aalto2013}.}
  \label{fig:VT}
\end{figure}

We use static MRI images acquired with a Siemens Magnetom Avanto 1.5 T scanner. A 12-element Head Matrix Coil, and a 4-element Neck Matrix Coil, allow for the Generalize Auto-calibrating Partially Parallel Acquisition (GRAPPA) acceleration technique. One speaker, a 26-year-old male, was imaged while he uttered four sustained Finnish vowels. The MRI data covers the vocal and nasal tracts, from the lips and nostrils to the beginning of the trachea, in 44 sagittal slices, with an in-plane resolution of 1.9mm. Figure \ref{fig:VT} shows the VT surface geometries extracted from MRI data using an automatized segmentation method \citep{Aalto2013}.

For our 1D acoustic analysis, we describe the vocal tract by an area function $A(x,t)$ where $0 \leq x \leq L$ is the distance from the glottis on the tube axis and $t$ denotes the time. We take the similar notion of \cite{Doel2008} in defining the variables $u(x,t)=A(x,t)\hat{u}/c$ and $p(x,t)=\hat{\rho}/\rho_{0}-1$ as the scaled versions of volume-velocity $\hat{u}$ and air density $\hat{\rho}$ respectively. $\rho_{0}$ is the mass density of the air and $c$ is the speed of sound. We solve for $u(x,t)$ and $p(x,t)$ in the tube using derivations of the linearised Navier-Stokes equation (\ref{eq:Navier}) and the equation of continuity (\ref{eq:cont}) subject to the boundary conditions described in equation \ref{eq:BoundaryCond}:
\small
\begin{subequations}
\begin{align} 
\tipasafemode \frac {\partial (u/A)}{\partial t}&+c \frac{\partial p}{\partial x}= -d(A)u + D(A)\frac{\partial^2 u}{\partial x^2} \label{eq:Navier} \\
\tipasafemode \frac {\partial (A p)}{\partial t}&+c \frac{\partial u}{\partial x}=- \frac{\partial A}{\partial t} \label{eq:cont} \\
u(0,t)&= u_{g}(t),\quad p(L,t)=0 \label{eq:BoundaryCond}
\end{align}
\label{Eq1}
\end{subequations}
\normalsize
\noindent where $d(A)= d_{0} A^{-3/2}$ and $D(A)= D_{0} A^{-3/2}$ with the wall loss coefficient $d_{0}=1.6$ $ms^{-1}$ and $D_{0}=0.002$ $m^3s^{-1}$; and $u_{g}(t)$ is the source volume velocity at the glottis. We couple the vocal tract to a two-mass glottal model \citep{Ishizaka1972} and solve equation \ref{Eq1} in the frequency domain using a digital ladder filter defined based on the cross-sectional areas of 20 segments of the vocal tract. We refer to \cite{Doel2008} for full details of the implementation.

For our 3D acoustic analysis, we calculate the vowel formants directly from the wave equation by finding the eigenvalues, $\lambda$, and their corresponding velocity potential eigenfunction, $\Phi_{\lambda}$, from the Helmholtz resonance problem:
\small
\begin{subequations}
\begin{align}
\tipasafemode \lambda^{2}\Phi_{\lambda}-c^{2}\Delta\Phi_{\lambda}=0&\qquad\text{on }\Omega\\
\tipasafemode \Phi_{\lambda}=0 &\qquad\text{on }\Gamma_{1}\\
\tipasafemode \alpha \lambda \Phi_{\lambda}+\frac{\partial \Phi_{\lambda}}{\partial \nu}=0 &\qquad\text{on } \Gamma_{2}\\
\tipasafemode \lambda \Phi_{\lambda} + c\frac{\partial \Phi_{\lambda}}{\partial \nu}=0 &\qquad\text{on } \Gamma_{3}
\end{align}\label{Eq2}
\end{subequations}
\normalsize
where $\Omega \in \mathbb{R}^3$ is the air column volume and $\partial \Omega$ is its surface including the boundary at mouth opening ($\Gamma_{1}$), at air-tissue interface ($\Gamma_{2}$) and at a virtual plane above glottis ($\Gamma_{3}$); and $\frac{\partial \Phi_{\lambda}}{\partial \nu}$ denotes the exterior normal derivative.The value of $\alpha$ regulates the energy dissipation through tissue walls, and the case $\alpha = 0$ corresponds with hard, reflecting boundaries. We calculate the numerical solution of equation \ref{Eq2} by Finite Element Method (FEM) using piecewise linear shape functions and approximately $10^5$ tetrahedral elements. The imaginary parts of the first two smallest eigenvalues $\lambda_{1}$ and $\lambda_{2}$ give first two Helmholtz resonances of the vocal tract. We refer to \cite{Aalto2014} and \cite{Kivela2013} for details of implementation.

In order to distinguish the effects of dimensionality (1D vs. 3D) from the effects of different boundary conditions in equations \ref{Eq1} and \ref{Eq2}, we also compute the Webster resonances by interpreting equation \ref{Eq2} in one dimension:

\small
\begin{subequations}
\begin{align}
\tipasafemode (\frac{\lambda^{2}}{c^{2}} \frac{1}{\Sigma^{2}}+\lambda\frac{2\pi \alpha W}{A})\Phi_{\lambda}&=\frac{1}{A}\frac{\partial}{\partial s}(A \frac{\partial \Phi_{\lambda}}{\partial s})&\text{on }[0,L]\\
\tipasafemode \lambda\Phi_{\lambda}-c\Phi_{\lambda}^{'}&=0 &\text{at } s=0\\
\tipasafemode \Phi_{\lambda}&=0 &\text{at } s=L
\end{align}\label{Eq3}
\end{subequations}
\normalsize
where $\Sigma$ denotes the sound speed correction factor that depends on the curvature of the vocal tract; A(x) is the area function and s is the implicit parameter to $\Phi_{\lambda}$, A, W and $\Sigma$. We refer to \cite{Kivela2015} for details of implementation and parameter values.

\section{Results and Discussion}

\begin{figure}[t]
\centering
 \captionsetup{font=small}
  \includegraphics[width=0.48\textwidth,keepaspectratio]{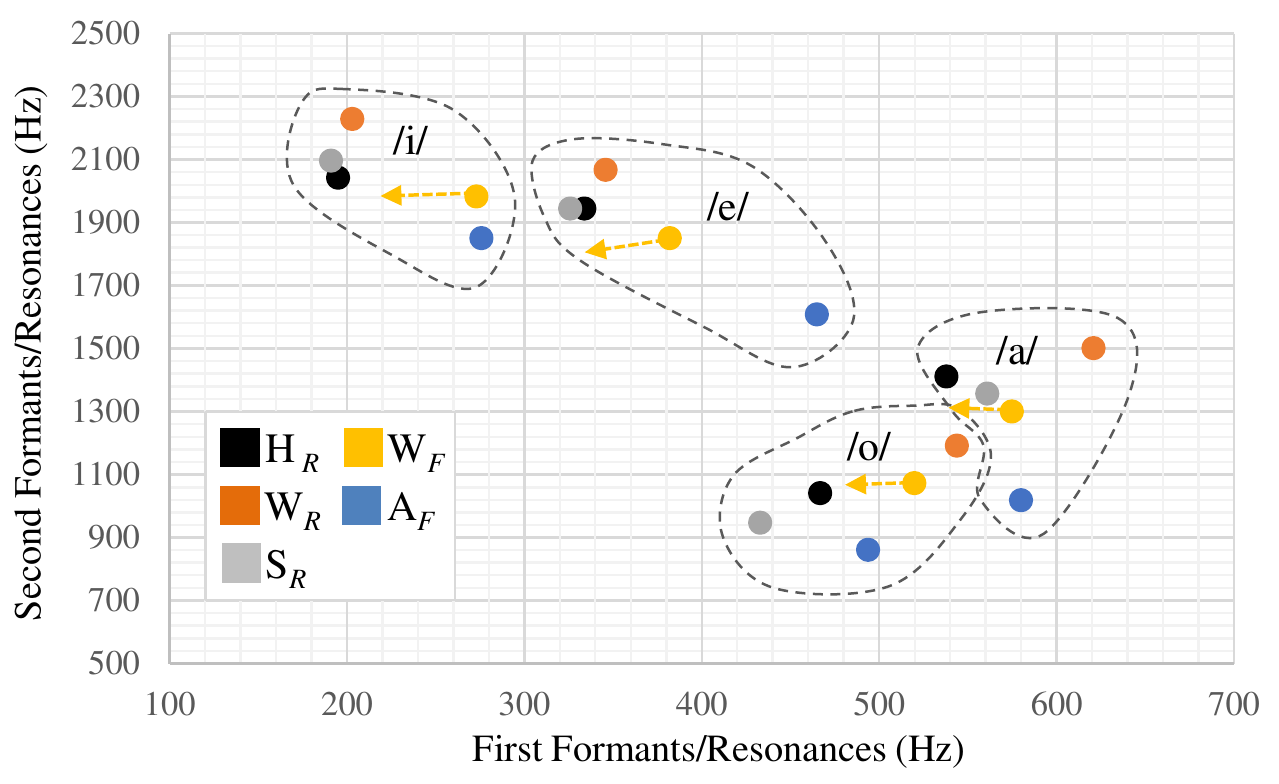}
 \caption [Simulation result for 1$^{\text{st}}$ and 2$^{\text{nd}}$ formant frequencies] {Simulation results for first and second formant/resonance frequencies for different vowels: Helmholtz resonances (H$_{R}$), Webster resonances (W$_{R}$) and their scaled version (S$_{R}$), Webster formants (W$_{F}$) and formants from audio signal (A$_{F}$). }
 \label{fig:twoformants}
\end{figure}

Figure \ref{fig:twoformants} shows the first two formant/resonance frequencies, computed for the four Finnish vowels. Webster formants (W$_{F}$) are calculated by solving Equation \ref{Eq1}, as suggested by \cite{Doel2008}. Helmholtz (H$_{R}$) and Webster resonances (W$_{R}$) are obtained from equations \ref{Eq2} and \ref{Eq3}, respectively \citep{Aalto2014}. S$_{R}$ denotes the scaled version of W$_{R}$. The figure also includes the formant frequencies (A$_{F}$) computed from audio signals recorded in an anechoic chamber \citep{Aalto2014}. The values are averaged over 10 repetitions of each vowel utterance.

As we can see in Figure \ref{fig:twoformants}, the resonance values (H$_{R}$, W$_{R}$ and S$_{R}$) lie close together for vowels /i/ and /e/, with S$_{R}$ being closer to H$_{R}$, as expected. For vowels /o/ and /a/ there is more difference in the first resonances of H$_{R}$ and W$_{R}$; For /o/, although S$_{R}$ lies closer to H$_{R}$, its first resonance is surprisingly low. For all of the vowels in Figure \ref{fig:twoformants}, the second formant of the audio is less than the computed results. The vowel /i/ is expected to be very sensitive to glottal end position, which, in turn, suggests the significance of adequate MRI resolution and accurate geometry processing for its spectral analysis.

Interestingly, the Webster formants (W$_{F}$) remain closer to the audio formants (A$_{F}$) than any of the resonances in the case of /i/, /e/, and /a/. For /o/ the distance to the A$_{F}$ is almost equal for W$_{F}$ and H$_{R}$, with both having similar values for the second formant/resonance; however, the first H$_{R}$ is lower, and the first W$_{F}$ is higher, than the first A$_{F}$.

The time-domain Webster analysis \citep{Doel2008} accounts for the VT wall-vibration phenomenon that is missing in the resonance analysis. This is done by substituting $A(x,t)$, from equation 5.3, with $A(x,t)+C(x,t)y(x,t)$: where $C(x,t)$ is the slow-varying circumference and $y(x,t)$ is the wall displacement governed by a damped mass-spring system. Setting $y(x,t)$ to zero, the Webster formants move along the arrows in Figure \ref{fig:twoformants}, reducing in their first formants. This moves the W$_{F}$ closer to the H$_{R}$ as both acoustical models now ignore the wall vibration. Meanwhile, W$_{F}$ moves away from the audio formants in the case of /i/, /e/, and /a/. The distance between W$_{R}$ and W$_{F}$ remains large, despite the fact that both acoustical models solve the Webster equation. The results imply that 3D Helmholtz analysis is more realistic than its 1D Webster version, as expected. 

Overall, our experiments suggest that the time-domain interpretation of acoustic equations provides more realistic results -- even if it requires reducing from 3D to 1D. This may be partially due to the fact that time-domain analysis allows for more complexity in the acoustical model such as inclusion of lip radiation and wall loss. Certainly unknown parameters always remain (such as those involved in glottal flow, coupling between fluid mechanics and acoustical analysis, etc.), which are estimated indirectly, based on observed behaviour in simulations.

It should be noted that our experiments are solely based on data from a single speaker. A larger database -- inclusive of more speakers from different genders and languages -- is needed in order to confirm the validity/generality of our findings.


\begin{thebibliography}{}
\small

\bibitem[Aalto \etal(2014)]{Aalto2014}
Aalto D, et al. 2014. Large scale data acquisition of simultaneous MRI and speech. J Appl Acoust. 83:64--75.


\bibitem[Aalto \etal(2013)]{Aalto2013}
Aalto D, et al. 2013. Algorithmic Surface Extraction from MRI Data-Modelling the Human Vocal Tract. Proceeding of 6th International Joint Conference on Biomedical Engineering Systems and Technologies; Barcelona, Spain.

\bibitem[Aalto \etal(2012)]{Aalto2012}
Aalto D, et al. 2012. How far are vowel formants from computed vocal tract resonances? arXiv:1208.5963.

\bibitem[Arnela and Gausch(2014)]{Arnela2014}
Arnela M, Guasch O. 2014. Three-dimensional behavior in the numerical generation of vowels using tuned two-dimensional vocal tracts. Proceeding of 7th Forum Acousticum; Kraków, Poland.


\bibitem[Doel and Ascher(2008)] {Doel2008} 
Doel K van den, Ascher UM. 2008. Real-time numerical solution of Webster's equation on a non-uniform grid. IEEE Trans Audio Speech Lang Processing 16:1163--1172.

\bibitem[Ishizaka and Flanigan(1972)]{Ishizaka1972} 
Ishizaka K, Flanigan JL. 1972. Synthesis of voiced sounds from a two-mass model of the vocal cords. J Bell Syst Tech. 51: 1233–1268.

\bibitem[Kivel\"a (2015)]{Kivela2015}
Kivel\"a A. 2015. Acoustics of the vocal tract: MR image segmentation for modelling, Master's thesis, Aalto University School of Science.

\bibitem[Kivel\"a \etal(2013)]{Kivela2013}
Kivel\"a A, Kuortti J, Malinen J. 2013. Resonances and mode shapes of the human vocal tract during vowel production. Proceedings of 26th nordic seminar on computational mechanics; Oslo, Norway.

\bibitem[Takemoto \etal(2014)]{Takemoto2014}
Takemoto H, Mokhtari P, Kitamura T. 2014. Comparison of vocal tract transfer functions calculated using one-dimensional and three-dimensional acoustic simulation methods. Proceeding of 15th Annual Conference of the International Speech Communication Association; Singapore, Singapore.

\end{thebibliography}
\end{document}